\documentclass[reqno,10pt]{amsart}
\usepackage{a4wide}
\usepackage[utf8]{inputenc}
\usepackage[english]{babel}
\usepackage{amsmath}
\usepackage{graphics}
\usepackage[dvips]{graphicx}
\usepackage{amsfonts}
\usepackage{rotating}
\usepackage{amssymb}
\usepackage{wasysym}

\renewcommand{\a}{\alpha}
\renewcommand{\t}{\tau}
\renewcommand{\b}{\beta}
\newcommand{\N}{\mathbb{N}}

\newtheorem{teorema}{Teorema}[section]
\newtheorem{teor}[teorema]{Theorem}
\newtheorem{prop}[teorema]{Proposition}
\newtheorem{oq}[teorema]{Open Question}
\newtheorem{lema}[teorema]{Lemma}
\newtheorem{defn}[teorema]{Definition}
\newtheorem{counter}[teorema]{Counterexample}

\title[Causality on Carter Spacetime]{Causal behaviour on Carter Spacetime}
\author[OF Blanco]{Oihane F. Blanco}
\address{Departamento de Matem\'aticas.
 Facultad de Ciencias, Colegio Polit\'ecnico, Universidad San Francisco de Quito.
Avda. Pampite y Orellana, Cumbay\'a, Quito, Ecuador}
\email{ofernandez@usfq.edu.ec}
\author[A moreira]{Andrea Moreira}
\address{Departamento de Matem\'aticas.
 Facultad de Ciencias, Colegio Polit\'ecnico, Universidad San Francisco de Quito.
Avda. Pampite y Orellana, Cumbay\'a, Quito, Ecuador}
\email{amoreira@usfq.edu.ec}

\thanks{Our sincere acknowledgment goes to Pr. J.M.M. Senovilla for his input and constructive feedback in the writing of this paper.}


\begin{document}

\begin{abstract}
 In this work we will focus on the causal character of Carter Spacetime (see \cite{carter},\cite{HE}). The importance of this spacetime is the following:  for the causally best well behaved spacetimes (the globally hyperbolic ones), there are several characterizations or alternative definitions. In some cases, it has been shown that some of the causal properties required in these characterizations can be weakened. But  Carter spacetime provides a counterexample for an impossible relaxation in one of them. We studied the possibility of Carter spacetime to be a counterexample for impossible lessening in another characterization, based on the previous results. 
 
 In particular, we will prove that the time-separation or Lorentzian distance between two chosen points in Carter spacetime is infinite. Although this spacetime turned out not to be the counterexample we were looking for, the found result is interesting {\it per se} and provides ideas for alternate approaches to the possibility of weakening the mentioned characterization.\\

\noindent\textsc{Resumen.} En esta investigación nos enfocamos en el carácter causal del espaciotiempo de Carter(ver \cite{carter},\cite{HE}). Este espaciotiempo es importante por la siguiente razón: para los espaciotiempos con un comportamiento causal óptimo, es decir, los globalmente hiperbólicos, existen varias caracterizaciones o definiciones alternativas. En algunos casos se ha demostrado que ciertas condiciones de causalidad requeridas en tales caracterizaciones pueden relajarse. Pero el espaciotiempo de Carter nos da un contraejemplo que hace imposible la relajación en una de ellas. Basándonos en estos resultados previos, estudiamos la posibilidad de que el espaciotiempo de Carter sea también un contraejemplo para otra caracterización.

En particular, demostraremos que la separación temporal o distancia Lorentziana entre dos puntos del espaciotiempo de Carter es infinita. Si bien este espaciotiempo resultó no ser el contraejemplo buscado, la conclusión es de por sí interesante y aporta  ideas alternativas para estudiar la posibilidad o no de rebajar la condición en la caracterización mencionada.
\end{abstract}
\maketitle

\section{Introduction}
 
In the intersection between General Relativity and Lorentzian Geometry there is an interesting theory, called the Theory of Causality, which studies  the causal relations between points 
 by analyzing the behaviour of the causal curves in a spacetime both locally and globally. We will focus our attention on the causal structure of Carter spacetime (see \cite{carter},\cite{HE}) and on the alternate characterizations for the causally best well-behaved spacetimes in the causal ladder: the globally hyperbolic spacetimes.
 
 In this section, we will  introduce some basic notation and conventions in Lorentzian geometry and an outline of this paper. We will not state any of the definitions and properties in Theory of Causality since, if interested, one can find an excellent review on the last advances on Theory of Causality in \cite{MS}, where all the required definitions, the causal conditions and the interplay between them are carefully studied. It is worth mentioning  that a new and interesting ordering of spacetimes has been recently carried out by García-Parrado and Senovilla in \cite{ SG3},\cite{SG}, and refined by García-Parrado and Sánchez in \cite{SG2}.

\subsection*{Notation and conventions}
In this article, a {\em spacetime} $(M, g)$ is a Haussdorf, paracompact, (space) orientable, time orientable, connected and differentiable manifold of dimension $n\geq 2$, together with a non-degenerate 2-covariant tensor field $g$ of signature\footnote {Depending on the convention, sometimes the signature $(-+{\ldots} +)$ is used instead.} $(+-{\ldots} -)$. A point $p$ in a spacetime is usually called {\em an event}. 
 The {\it Lorentzian metric} $g$ over a spacetime ensures the existence of a unique torsion-free ($\nabla_x Y -\nabla_Y X=[X,Y]$) connection or covariant derivative, the {\it Levi-Civita connection $\nabla$} on $M$, under which the metric $g$ is covariantly constant ($\nabla g=0$). It also allows to define the causal character of a  vector $v\in T_p M$: it is {\it timelike} if $g_p(v,v)>0$; {\it spacelike} if $g_p(v,v)<0$; {\it lightlike} if $g_p(v,v)=0$ and $v\neq 0$; {\it causal} if it is timelike or lightlike, and {\it null} \footnote{Sometimes only the zero vector is called null.} if it is lightlike or the zero vector. At each point $p$ of a {\it Lorentzian manifold} $(M,g)$, i.e. a differentiable manifold endowed with a Lorentzian metric, the set of all causal vectors in the tangent space $T_p M$ is called the {\it causal cone} over $p$. This cone has two connected components. A time orientation at $p$ is the choice of one of the connected components, which will be called {\it the future causal cone}; obviously, the other component is called the {\it past causal cone}\footnote{In a similar way, one can define future/past timelike cones, and future/past lightlike cones.}.  
 Moreover, a spacetime is {\it time orientable} if and only if there exists a globally defined (non-unique) timelike vector field $V$, which will give the future direction of the spacetime. In that case, any causal vector $v\in T_p M$ is {\it future directed} if and only if $g_p(v,V_p)>0$ and {\it past directed} if and only if $g_p(v,V_p)<0$. 
  The causal cones defined on a spacetime $M$ are a subset of $T_p M$ at each point $p$. Recall that globally the metric of a spacetime changes from point to point and that causes the cones to twist when moving through it.
  
  Let $\alpha$ be a differentiable curve with $d\a\neq 0$ (i.e., with non-vanishing tangent vector field $\a '$) defined on an interval $I\subset \mathbb{R}$, being $\tau\in I$ the parameter of the curve. The end values $a$ and $b$  of the interval  may be infinity, that is  $-\infty\leq a<b\leq \infty$.  The curve $\a$ is {\it timelike, lightlike, spacelike} or {\it causal} if its tangent vector $\a'$ is timelike, lightlike, spacelike or causal respectively, at every $\tau\in I$.  Let $p,q$ be two points in $M$. We say that $\alpha$ {\it connects} $p$ with $q$ if $\a(a)=p$ and $\a(b)=q$. Moreover, a causal curve $\a$ is {\it future directed} if its tangent vector is future directed at every $\tau\in I$. Let now $\a$ be a future directed causal curve.  If $\displaystyle\lim_{\t\rightarrow a}\a(\tau)=p$, the event $p$ is the {\it past endpoint} of the curve, and if $\displaystyle\lim_{\t\rightarrow b}\a(\tau)=q$, the event $q$ is the {\it future endpoint} of the curve. If there is no future/past endpoint, the curve is said to be {\it future/past inextendible}; if there are no endpoints, the curve is said to be {\it inextendible}.
  A piecewise differentiable curve $\a'$ is said to be {\it timelike, lightlike, spacelike} or {\it causal} if in each differentiable piece $\a'$ is a  timelike, lightlike, spacelike or causal curve respectively, and the two lateral tangent vectors at each break lie in the same causal cone. From now on, if it is not specified, a curve will be piecewise differentiable. 
The {\it length} over a spacetime of a piecewise differentiable timelike, lightlike or spacelike curve $\a(\t)$ between two points $p=\a(a)$ and $q=\a(b)$ is defined as:
\begin{equation}\label{AL}
L(\a)=\displaystyle\int_a^b \sqrt{|g(\a'(\t),\a'(\t))|}d\t.
\end{equation}

Let $C(p,q)$ be the set of all future directed piecewise differentiable causal curves connecting $p$ with $q$. The {\it Lorentzian distance} or {\it time separation between the points $p$ and $q$} on a Lorentzian manifold $(M,g)$ is defined as:
$$d(p,q)=\displaystyle \sup_{\a\in C(p,q)}L(\a)$$

Let now $(M,g)$ be a spacetime and $p\in M$. The {\em chronological future of $p$}, denoted by $I^+(p)$, is defined as the set of points in $M$ that can be connected with $p$ by a (piecewise differentiable) past directed timelike curve. The {\em causal future of  $p$}, denoted by  $J^+(p)$, is defined as the set of points in $M$ that can be connected with $p$ by a (piecewise differentiable) past directed causal curve, plus $p$.
The chronological past $I^-(p)$ and causal past $J^-(p)$ are defined in an analogous way. 
There is another interesting set, used in the next sections, which is nothing but the intersection of the causal future of $p$ and the causal past of $q$: $J(p,q)=J^+(p)\cap J^-(q).$

We shall also mention the Lorentzian structure inherited by a submanifold $M'$ of a Lorentzian manifold $(M,g)$. There is a classification similar to that of the vectors of $T_p M$. If $g_{|M'}$ is the Lorentzian metric restricted to vector fields in $TM'$, then $M'$ is a {\it timelike submanifold} if $g_{|M'}$ is a Lorentzian metric; a {\it spacelike submanifold} if $g_{|M'}$ is a Riemannian metric, and  a {\it lightlike submanifold}  if the bilinear form $g_{|M'}$ is degenerate.

  \subsection*{Outline of this paper}
 
The spacetime we are focusing on in this article is defined in \cite{carter}, and we will refer to it as Carter spacetime. Our interest on Carter spacetime arose from the fact that it was used as a counterexample, proving that the strong causality condition in a spacetime cannot be weakened at the same time as the compactness condition, if one were to characterize globally hyperbolic spacetimes (see counterexample \ref{co}). For that reason, a complete section is left to an analysis of Carter spacetime and its causal behaviour (Section \ref{carter}). In that section we will prove that Carter spacetime is causal (section \ref{cau}) but not strongly causal (section \ref{strong}), that the set $\overline{J(p,q)}$ is compact (section \ref{jpq}) and, the main result of this article,  that the Lorentzian distance between some points is infinite (section \ref{dist}, Proposition \ref{proo}).  The global hyperbolicity condition, which is the strongest in the causal ladder, is crucial in the Singularity theorems of General Relativity, and its various equivalent definitions are the motivation of this work. For that reason, it will be addressed in more detail in section \ref{gh}. Finally, as a consequence of our result, we will discuss some points and open questions related to the subject of global hyperbolicity in section \ref{oq}.

\section{Carter Spacetime}\label{carter}

The differentiable manifold $M$ on this particular spacetime is defined as a quotient space of $\mathbb{R}^{3}$ diffeomorphic to $\mathbb{R}\times\mathbb{T}^2$, with the following identifications:	
	\begin{eqnarray*}(t,y,z)&\sim{ }_1&(t,y, z+1),\\
	(t,y,z)& \sim{ }_2& (t, y+1,z+a), \text{ with } {a\in(0,1)\cap (\mathbb{R}\setminus\mathbb{Q})}.\end{eqnarray*}
	\par
\begin{figure}
\begin{center}
\includegraphics[scale=1.5]{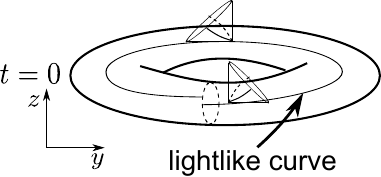}
\end{center}
\caption{The lightlike hypersurface $S_0=\{t=0\}$ in Carter Spacetime.}
\label{toricarter}
\end{figure}

 For every surface $\{t=t_0\}$ with $t_0$ constant, we obtain a cylinder under the first equivalence relation, and under the second equivalence relation we get a torus in which the points $(0,z)$ are identified to $(1,z+a)$. This last set of equivalence classes will be crucial to prove that the spacetime is causal since, when identifying the edges of the cylinder after the first identification, the torus twists a little, so that the closed causal curve keeps turning around it, without ever closing, as it will be shown in the next section (see figure \ref{toricarter}). 

The metric in Carter spacetime is defined by (see \cite{carter},\cite{HE}):
\begin{equation*}\label{Carter}
{ds^2=c(t)\left( dt^2-dy^2\right)+ 2 dt dy-dz^2,}
\end{equation*}
	with $c(t)=(\cosh (t)-1)^2$. The signature for this metric is $(1,-1,-1)$ and the time orientation is such that the causal vector field  $\partial_t$ is directed to the future. Notice that, if $t_0\neq 0$, the surface $S =\{ t=t_0\}$ is spacelike, since the metric there becomes 
$$g_{|S}=-c(t_0)dy^2-dz^2,$$
with signature $(-,-)$. By contrast, in the surface $S_0=\{ t=0\}$  the metric is degenerate:
\begin{equation*}\label{de}
	g_{|S_0}= -dz^2
	\end{equation*}
and the surface is lightlike.
\subsection{Carter spacetime is causal}\label{cau}
Our aim is to prove that the lightlike curve in figure \ref{toricarter} is an endless non-closed causal curve, as it is suggested in the figure. Let $t_0\neq 0$ be a fixed value. Since all vectors lying in $S=\{t=t_0\}$ are spacelike, the causal cone of any point $p$ in $S$ does not intersect $S$. This means that the tangent vector of any future directed causal curve passing through $S$ must have an increasing $t$-coordinate, and thus the curve cannot be closed unless it reaches $t=0$. Therefore, we can restrict our search to the surface $S_0=\{t=0\}$, which is a lightlike surface as seen above. 

For a curve $\alpha(\tau)=(0,\alpha_2(\tau),\alpha_3(\tau))$ in $S_0$ to be causal, it must satisfy the condition
	$$g(\alpha'(\tau),\alpha'(\tau))=-(\alpha_3'(\tau))^2\geq 0 $$ and therefore  $\alpha_3'(\tau)=0$.
Consequently, up to reparametrizations, the causal curves that could be closed on the surface $S_0$ would be of the form:
	$$\alpha(\tau)=(0,\alpha_2(\tau),k).$$
	
 Let us prove that the curve $\alpha(\tau)=(0,\alpha_2(\tau),0)$ starting at $(0,0,0)$ is not closed ---the others can be obtained by a translation of this one. Assuming that $\alpha$ is a future directed piecewise differentiable curve, that is $ g(\alpha ',\partial t)>0$, one concludes that $\alpha_2'>0$. A convenient reparametrization leads to $\alpha(y)=(0,y,0)$. 
 
By the definition of Carter spacetime, the event $(0,0,0)$ has the following equivalent events:
\begin{center}
$(0,0,0)\sim_1{ }(0,0,n)\sim_2 (0,m,n+am)$, with $n,m\in \mathbb{Z}-\{0\}$.
\end{center}
For the curve $\alpha$ to be closed, there must exist a parameter $y_0\neq 0$ such that  $\alpha (y_0)$ is in the equivalence class of $(0,0,0)$. That is,  $(0,y_0,0)=(0,m,n+am)$, for some $m,\,n \in \mathbb{Z}$. But this is impossible since $m=y_0\neq 0$ and $a$ is irrational.
	
	\subsection{Carter spacetime is not strongly causal}\label{strong}
One of the consequences of being a strongly causal spacetime is that there cannot exist totally or partially imprisoned causal curves in a compact set (see for instance \cite[Prop. 3.13]{MS}). A causal curve is totally (resp. partially) imprisoned in a compact set if once it enters the set it never leaves it (resp. if the curve leaves the set, it will continually re-enter it). But the causal curve constructed in section \ref{cau} is contained in $\{0\}\times\mathbb{T}^2$, which is a compact set. Therefore Carter spacetime is not strongly causal.

\subsection{The set $\overline{J(p,q)}$ is compact in Carter spacetime}\label{jpq}
Let $p,q$ be two points in Carter spacetime and $t_p$, $t_q$ the $t$-coordinates of $p$ and $q$, respectively.  Since any future directed causal curve connecting $p$ and $q$ must have  a non-decreasing $t$-coordinate, we may conclude that  $J(p,q)\subset [t_p,t_q]\times \mathbb{T}^2$, which is compact. Therefore the closure  $\overline{J(p,q)}$ is compact\footnote{Every closed subset of a compact set is also compact.}, for all $ p,q$ in $M$. This interesting property  was pointed out for the first time  in \cite{Ga}. It makes of Carter spacetime a counterexample to the question of whether the conditions of strong causality and compactness of $J(p,q)$ for a globally hyperbolic space could be simultaneously replaced by the weaker conditions of causality and compactness of $\overline{J(p,q)}$ (see Section \ref{gh}).

\subsection{The time-separation in Carter spacetime reaches infinite value}\label{dist}
The main result of this paper is presented in the following proposition:
\begin{prop}\label{proo}
There exist at least two points $p$ and $q$ in Carter spacetime such that their Lorentzian distance $d(p,q)$ is infinite. 
\end{prop}

The arc length functional given in formula \eqref{AL} for causal curves $\alpha(\tau)=(t(\tau),y(\tau),z(\tau))$ in Carter spacetime becomes:
\begin{equation}\label{arc}
L(\alpha)=\int_{\tau_0}^{\tau_1}\sqrt{g(\alpha',\alpha')}d\tau=\int_{\tau_0}^{\tau_1}\sqrt{c(t(\tau))({t'(\tau)}^2-{y'(\tau)}^2)+2 y'(\tau)t'(\tau)-{z'(\tau)}^2}d\tau.
\end{equation}

Our goal is to build a sequence of timelike curves $\{\alpha_n\}_{n\in\mathbb{N}}$ connecting a point  $p$ in $\{t=t_p<0\}$ to another point $q$ in $\{t=t_q>0\}$ in such a way that the lengths of those curves are not bounded from above. Since Carter spacetime was obtained by the projection $\Pi:\mathbb{R}^3\longrightarrow\mathbb{R}\times\mathbb{T}^2$, the elements of  $\mathbb{R}^3$ will be denoted with a $\hspace{0.2cm}\sim\hspace{0.2cm}$ above them and their images under the projection without it. In order to prove Proposition \ref{proo} we need some auxiliary lemmas:
\begin{lema}\label{lem1}
In Carter Spacetime, for each $p$ with $t_p<0$, there exists a future inextendible timelike curve ${\beta}$ starting at $p$ and never intersecting $S_0=\{t=0\}$,  such that $L(\beta)=\infty$. 
\end{lema}

{\it Proof.}

 Let $p=(t_p,y_p,z_p)$ be a point in Carter spacetime, with $z_p$ a value in the interval $[0,1)$ and $t_p<0$. The length functional \eqref{arc} over the (not yet projected) curves\footnote{since $\tilde{z}_p=z_p$ and $\tilde{t}=t$  for any $t$, we will not use $\hspace{0.2cm}\sim\hspace{0.2cm}$.} in $\mathbb{R}^3$ given by $\tilde{\alpha}(\tau)=(t(\tau),\tilde{y}(\tau),z_p)$, after a reparametrization of the type $t=t(\tau)$ becomes:
$$ L\left(\tilde{\alpha}\right)=\int_{t_p}^{t_1}\sqrt{c(t)(1-{\tilde{y}'(t)}^2)+2 \tilde{y}'(t)}dt. $$
 Let $t_1=0$. For any $t\in(t_p,0)$, the radicand in $L$ is a quadratic polynomial in  $\tilde{y}'(t)$ that reaches its maximum value at $\tilde{y_0}'(t)=\frac{1}{c(t)} $, thus
\begin{equation}\label{y}
\tilde{y}_0(t)=\frac{\sinh(t)(\cosh(t)-2)}{3 c(t)}+k.
\end{equation}
Take $k\in\mathbb{R}$ such that $\tilde{y}_0(t_p)=y_p$. Note that $\tilde{y}_0(t)$ is defined for $t<0$.  Define over the interval $[t_p,0)$ the curve $\tilde{\beta}(t)=(t,\tilde{y}_0(t),z_p)$. This curve starts at $\tilde{p}$, which becomes $p$ in Carter spacetime,  and satisfies:
\begin{equation}\label{L1}
L\left(\tilde{\beta}\right)=\int_{t_p}^{0}\sqrt{\frac{1+c(t)^2}{c(t)}}dt>\int_{t_p}^{0}\sqrt{\frac{1}{c(t)}}dt=\infty .
\end{equation}
Since $c(t)$ is positive for any $t<0$, it results that $g(\tilde{\beta}',\tilde{\beta}')=g(\tilde{\beta}',\partial_t)=\frac{1+c(t)^2}{c(t)}>0$, therefore $\tilde{\beta}$ is a future directed timelike curve, and it is future inextendible because $\displaystyle \lim_{t\rightarrow 0}\tilde{y}_0(t)=\infty$ . Indeed, $t$ never reaches zero so that this curve does not intersect $S_0=\{t=0\}$.

\begin{figure}[htb]
\centering 
\includegraphics[width=8cm,height=6cm]{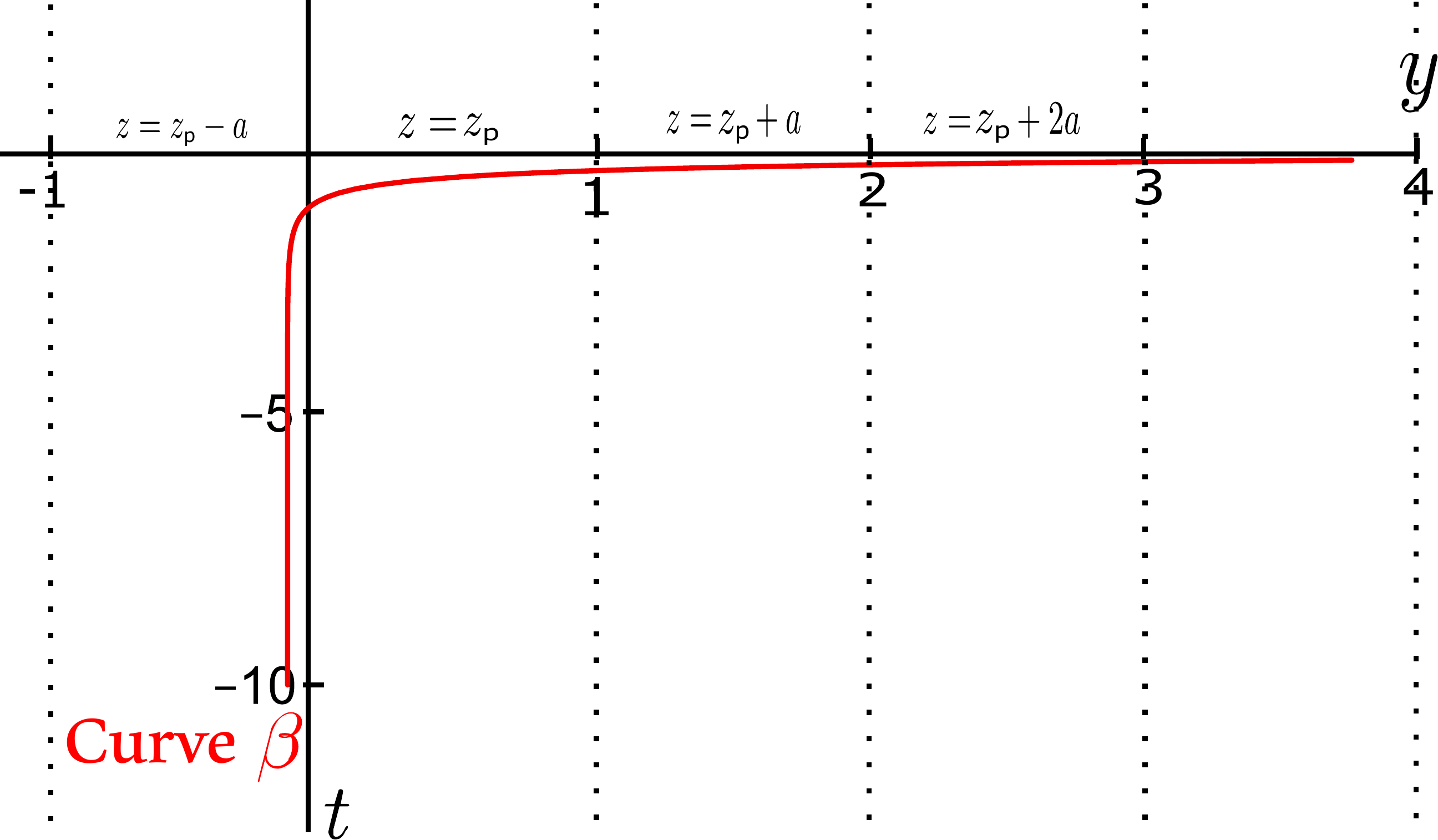}

\caption{\scriptsize The curve ${\beta}$ defined in $[t_p,0)$, in Carter spacetime.} 
\label{beta}
\end{figure}

 The required curve $\beta$ in Carter spacetime is obtained by projecting the curve $\tilde{\beta}$ (see figure \ref{beta}).
			
\begin{flushright}
$\square$
\end{flushright}

Recall that our aim is to construct a sequence of timelike curves $\{\alpha_n\}_{n\in\mathbb{N}}$ connecting a point  $p$ in $\{t=t_p<0\}$ to another point $q$ in $\{t=t_q>0\}$. The reasoning above suggests that the value of $L$ (for an appropriate choice of $z'=0$) cannot be bounded for a sequence of curves $\{{\alpha}_n\}_{n\in\mathbb{N}}$ built in the following way: if $\{t_n\}_{n\in\mathbb{N}}$ is a sequence of negative values converging to zero, then for each $n\in\mathbb{N}$ and over the interval $[t_p,t_n]$,  the curve $\tilde{\alpha}_n$ starting at $p$ is equal to $\tilde{\beta}$; for $t>t_n$, the curve $\tilde{\alpha}_n$ will be glued to another curve that after a while will reach $\{t=0\}$. It is important to note that, while connecting the points $p$ and $q$, due to the identifications of the spacetime (see figure \ref{beta}),  we cannot ignore de $z$-coordinate. That is the reason why we chose a point $q$ with $t_q>0$ instead of $t_q=0$.

\begin{lema}\label{lem2}
In Carter spacetime, there exists a
sequence of piecewise differentiable future directed timelike curves $\{\alpha_n\}_{n\in\mathbb{N}}$ starting at a point $p$ with $t_p=-1<0$ and endpoints in $S_0=\{t=0\}$, such that $\displaystyle \lim_{n\rightarrow \infty} L(\alpha_n)=\infty$.
\end{lema}

{\it Proof.}

Let $t_p=-1<0$ be the $t$-coordinate of $p$ and $\{t_n\}_{n\in\mathbb{N}}$ be a sequence of  values $-1\leq t_n\leq 0$ converging to zero. If $\tilde{y}_0(t)$ is the function given in \eqref{y}, define the sequence of functions $\tilde{y}_n(t)$ as:
$$\tilde{y}_n(t)=\begin{cases}
\tilde{y}_0(t)&\text{ if }-1\leq t<t_n,\\
A_n(t-t_n)+\tilde{y}_0(t_n)&\text{ if }t_n\leq t\leq 0.\end{cases}$$
\noindent for some $A_n$ to be determined.  For each $n\in\mathbb{N}$, the curve $\tilde{\alpha}_n(t)=(t,\tilde{y}_n(t),z_p)$ defined over $[-1,0]$ is, by construction, the curve $\tilde{\beta}$ in Lemma \ref{lem1},  up to  $t_n$. Thus, in $[-1,t_n)$ it is  piecewise differentiable and future directed timelike. Therefore, in the same interval, the projection $\alpha_n$ of each curve must also be a future directed timelike  curve in Carter spacetime. Moreover, $\tilde{\alpha}_n$ connects the point $\tilde{p}=\tilde{\beta}(-1)$ with $\tilde{q}_n=(0,\tilde{y}_n(0),z_p)$. Notice that  $\tilde{y}_n(t_n^{-})=\tilde{y}_n(t_n^{+})=\tilde{y}_0(t_n)$, and the final points of $\tilde{\alpha}_n$ and $\alpha_n$ do not coincide due to the identifications of points. Finally, since the $t$-coordinate increases, the curve, if timelike, automatically will be future directed\footnote{Because the projected vectors $\alpha_n'(t_n^-)$ and $\alpha_n'(t_n^+)$ both belong to the same timelike cone.} in the entire interval. 

Thus, it only remains to find $A_n$ such that, for each $n\in\N$, the curve ${\alpha}_n(t)$ is timelike over $[t_n,0]$. For ${\alpha}_n$ to be a timelike curve, the curve $\tilde{\alpha}_n$ must satisfy: 
		\begin{equation}\label{an}
		\tilde{g}(\tilde{\alpha}_n'(t),\tilde{\alpha}_n'(t))=c(t)(1-A_n^2)+2A_n>0\hspace{0.2cm},\forall t\in[t_n,0].
		\end{equation}
		It is easy to show that (\ref{an}) is true if:
$$ \frac{1-\sqrt{1+c(t)^2}}{c(t)}<A_n<\frac{1+\sqrt{1+c(t)^2}}{c(t)}, \hspace{0.5 cm} \forall t \in(t_n,0),\forall n\in\mathbb{N}. $$	
	Equivalently, if:
$$ \displaystyle \sup_{t\in[t_n,0)}\left(\frac{1-\sqrt{1+c^2(t)}}{c(t)}\right)<A_n<\displaystyle \inf_{t\in(t_n,0)}\left(\frac{1+\sqrt{1+c^2(t)}}{c(t)}\right), \hspace{0.5cm} \forall n\in\mathbb{N}.$$

\begin{figure}[htb]
\centering 
\includegraphics[width=6cm,height=5cm]{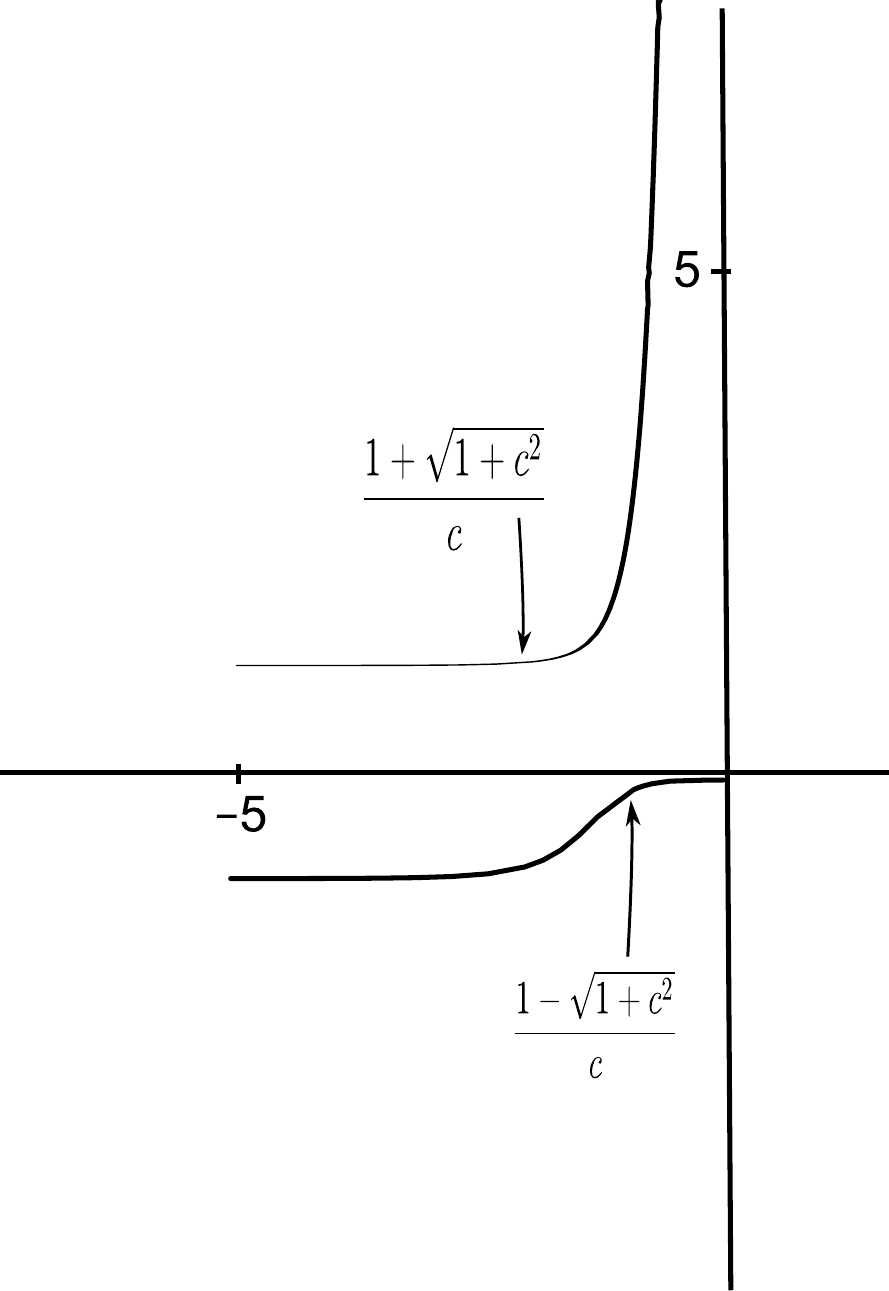}

\caption{\scriptsize Graph of the functions $\frac{1\pm\sqrt{1+c(t)^2}}{c(t)}$ for $t<0$.}
\label{fun}
\end{figure}
	 
	Clearly we have in one hand that $\frac{1-\sqrt{1+c^2}}{c}$ is always negative and its supremum is zero, and on the other hand, since $\frac{1+\sqrt{1+c(t)^2}}{c(t)}$ is increasing (see figure \ref{fun}),
	 $$\displaystyle \inf_{t\in(t_n,0)}\left(\frac{1+\sqrt{1+c(t)^2}}{c(t)}\right)=\frac{1+\sqrt{1+c(t_n)^2}}{c(t_n)}.$$
	   The function $c(t)$ is strictly positive for all $t<0$, thus $\frac{1+\sqrt{1+c(t_n)^2}}{c(t_n)}>\frac{2}{c(t_n)}$. Therefore, if we choose the constants $\{A_n\}_{n\in\mathbb{N}}$ verifying $0<A_n\leq D_n=\frac{2}{c(t_n)},\forall n\in\mathbb{N}$, the curve will always be timelike. For instance, we may choose: \\
$$A_n=\frac{\tilde{y}_0(t_n)-\left\lceil \tilde{y}_0(t_n)\right\rceil -1}{t_n},$$ 
where $\left\lceil x\right \rceil$ is the ceiling value\footnote{The closest integer greater or equal to $x$.} of $x$. For this particular case we may see that both the numerator and the denominator of $A_n$ are negative, hence $A_n>0$.

To prove that $A_n\leq D_n,\forall n\in\mathbb{N}$, we just need to check that $\dfrac{A_n}{D_n}<1$, but this is true since
$$
0\leq \dfrac{A_n}{D_n}=\dfrac{\left(\left\lceil \tilde{y}_0(t_n)\right\rceil-\tilde{y}_0(t_n)+1\right)c(t_n)}{2|t_n|}\leq\frac{c(t_n)}{|t_n|},$$
and 
$$\frac{c(t_n)}{|t_n|}<1,\hspace{0.5cm} \forall t_n\in [-1,0]. $$

Therefore $A_n\leq D_n,\forall n\in\mathbb{N}$. 

Finally, $L(\alpha_n)=L(\alpha_{n|[-1,t_n]})+L(\alpha_{n|[t_n,0]})$ and trivially $\displaystyle\lim_{n\rightarrow \infty}L(\alpha_{n|[t_n,0]})=0$. In consequence, 
$$\displaystyle\lim_{n\rightarrow\infty}L(\alpha_n)=\displaystyle\lim_{n\rightarrow\infty}\int_{-1}^{t_n}\sqrt{\frac{1+c(t)^2}{c(t)}}dt=L(\beta)=\infty.$$
\begin{flushright}
$\square$
\end{flushright}

Now we have the tools needed to prove the proposition:

{\it Proof of Proposition \ref{proo}}

\begin{figure}[htb]
\centering 
\includegraphics[scale=1]{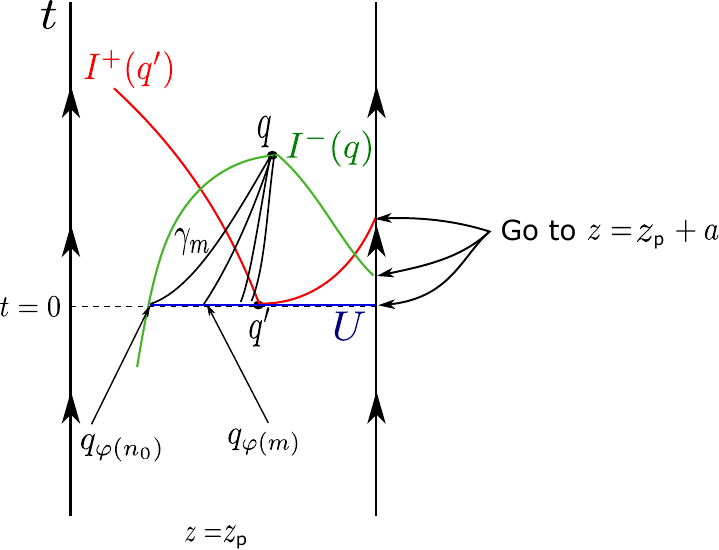}
\caption{\scriptsize The set $U$ and the curves $\gamma_m$.}
\label{prop}
\end{figure}

The projected sequence of curves $\{\alpha_n\}_{n\in\mathbb{N}}$ in Lemma \ref{lem2} over Carter spacetime leads to a sequence of points $\{q_n\}_{n\in\mathbb{N}}=\{\alpha_n(0)\}_{n\in\mathbb{N}}$ over the compact set $\{0\}\times\mathbb{T}^2$. Since any sequence of points in a compact metric space admits a converging subsequence, we can assume that there exists a subsequence $\{q_{\varphi(n)}\}_{n\in\mathbb{N}}$ of $\{{q}_n\}_{n\in\mathbb{N}}$ that converges to a point $q'$  in $\{0\}\times\mathbb{T}^2$. Let $q$ be a point in the chronological future $I^+(q')$ of $q'$. The intersection $U=I^-(q)\cap\left(\{0\}\times\mathbb{T}^2\right)$ is an open set of  $\{0\}\times\mathbb{T}^2$ containing $q'$ (see Figure \ref{prop}).
 As a consequence, there exists a natural number $n_0\in\mathbb{N}$ such that for all $ m$ greater than $n_0$, the points $q_{\varphi(m)}$ are contained in $U$. This implies the existence of a future directed timelike  curve $\gamma_m$  that connects $q_{\varphi(m)}$ and $q$  for each $m>n_0$. After a proper reparametrization of $\gamma_m$, we consider the sequence of future directed timelike curves resulting from appending the previous two: 
 $$\{\sigma_m\}_{m\in\mathbb{N},m>n_0}=\{\gamma_m*\alpha_{\varphi(m)}\}_{m\in\mathbb{N},m>n_0},$$
 that is,
  $$\sigma_m(t)=\begin{cases}\alpha_{\varphi(m)}(t,)&\text{ if }t\in[t_p,0],\\\gamma_m(t),&\text{ if }t\in(0,t_q].\end{cases}$$
  These curves connect $p$ with $q$ and using the inverted triangular inequality (e.g. see \cite[Lemma 14.16]{On}) we have:
$$p\ll q_{\varphi(m)}\ll q\Longrightarrow d(p,q)\geq d(p,q_{\varphi(m)})\geq L(\alpha_{\varphi(m)}), \forall m\in\mathbb{N}, m>n_0.$$
Finally, by lemma \ref{lem2},
\begin{eqnarray*}
d(p,q)&\geq&\displaystyle\lim_{m\rightarrow\infty}L(\alpha_{\varphi(m)})=\infty.
\end{eqnarray*}
\begin{flushright}
$\square$
\end{flushright}

\section{Characterizations of global hyperbolicity}\label{gh}

The notion of global hyperbolicity is crucial in General Relativity; the Einstein initial condition problem, the singularity theorems due to Hawking, Penrose or Gannon, the theorems related to the mass or to {\it the cosmic censor conjecture} are all formulated in global hyperbolic spacetimes ---or neighborhoods--- (see \cite{senovilla} for an exhaustive study of singularity theorems, and references therein). This name comes from the solution of the wave equation for a $\delta-$function source at a point $p$, since in such spacetimes this equation has unique solution. Moreover, if we restrict our definition to a globally hyperbolic neighborhood $N$, then outside $N-J^{+}(p)\cap N$ the solution vanishes \cite[chapter 7]{HE}. There exist various alternative definitions for global hyperbolicity, that have been adjusted in recent years. The more clarifying one,  due to Geroch, is presented in Definition \ref{def2}. It represents the realistic structure of a globally hyperbolic spacetime. As one can see in the definition, that structure is substantially simplified. The idea is that, in a globally hyperbolic spacetime, the knowledge of a hypersurface in an instant of time can determine all the spacetime:

\begin{defn}\label{def2}
A spacetime $(M,g)$ is globally hyperbolic if and only if it admits a Cauchy hypersurface $\Sigma$, that is, a topological hypersurface which is cut only once by each inextendible timelike curve. Then, $M$ is homeomorphic to the foliation $\mathbb{R}\times \Sigma$, and each $\{t_0\}\times \Sigma$ is a Cauchy hypersurface.
\end{defn}
An achronal set is a set in which there are no points chronologically connected ---i.e., $I^+(S)\cap S=\emptyset$. For instance, the set $S_0$ in Carter spacetime is achronal. The domain of dependence $D(S)$ of a closed achronal set $S$ is the union of the future and past domains of dependence, denoted by $D^+(S)$ and $D^-(S)$ respectively. These last two sets are defined as the sets of events satisfying that every future/past directed inextendible timelike curve through the event intersects $S$ (see \cite{Ge} for more details). For example, any signal sent to $D^+(S)$ must be registered in $S$, and given the appropriate information about initial conditions on $S$, one will be able to predict what happens at any point in  $D^+(S)$. Similar properties can be deduced for $D^{-}(S)$. Summarizing, $D(S)$ is the complete set of events for which all conditions should be determined by the knowledge of conditions on $S$.  Then, a Cauchy hypersurface $\Sigma$ can also be defined as a closed achronal set with $D(\Sigma)$ being all the spacetime. Thus, the entire future and past history of the universe can be predicted from conditions at the instant of time represented by $\Sigma$.  Even more, $\Sigma$ is a 3-dimensional, topological, closed, spacelike and $C^0$-submanifold which supplies information about an instant of time of the universe. Observe that the foliation provides a global time $f$ for which $f=constant$ is a Cauchy hypersurface. There are good reasons to believe that physically realistic spacetimes must be globally hyperbolic. But if not, one can always restrict the study to a globally hyperbolic neighborhood. 

However, the first definition for globally hyperbolic spacetimes was formulated in a completely different way by Leray in 1952 (see \cite{Le}). To introduce this definition, we need some previous concepts: let $C(p,q)$ be the set of piecewise differentiable future directed causal curves from $p$ to $q$, up to a reparametrization. Obviously, if $q\not\in J^+(p)$, then $C(p,q)=\emptyset$. We can define a topology over $C(p,q)$, called the {\it $C^0$-topology}, by taking the open sets as the union of sets of the type:
$$O(U)=\left\{\gamma\in C(p,q):\gamma\subset U\right\},$$
where $U$ is an open set in $M$. In other words, $O(U)$ consists on all causal curves form $p$ to $q$ which lie entirely within $U$. Now we are able to introduce the second definition due to Leray:
\begin{defn}\label{def1}
A spacetime $(M,g)$ is globally hyperbolic if and only if it is {\bf strongly causal} and, for each $p,q \in
M$ the space of causal curves that connect them is compact under the $C^0$-topology.
\end{defn}
The equivalence between Definition \ref{def2} and \ref{def1} was proven in a known theorem by Geroch (\cite{Ge}). Those definitions yield a third one:
\begin{defn}\label{def3}
A spacetime $(M,g)$ is globally hyperbolic if and only if it is {\bf strongly causal} and for any two points $p,q\in M$ the set $J(p,q)$  is compact.
\end{defn}
The equivalence between all these three definitions is provided  in \cite[pp. 205-209]{wald}. Aditionally, in \cite[lemma 4.29]{BE} it was proven that, in definition \ref{def3}, it is possible to simplify the compactness condition of $J(p,q)$, to compactness of its closure  $\overline{J(p,q)}$.

\begin{teor}\label{teo1}
A spacetime $(M,g)$ is globally hyperbolic if and only if it is  strongly causal and for any two points $p,q\in M$ the set ${\bf\overline{J(p,q)}}$  is compact.
\end{teor}
Under the conditions of Theorem \ref{teo1}, it results that $J(p,q)$ is closed. Moreover, the definition \ref{def3} has been  simplified  a little more in another way, proving that compactness of $J(p,q)$  plus causality implies strong causality, that is: 
\begin{teor} \cite{BS}
A spacetime $(M,g)$ is globally hyperbolic if and only if it is {\bf causal} and for any two points $p,q\in M$ the set $J(p,q)$  is compact.
\end{teor}

 The key in the proof of this theorem is that causality plus compactness of $J(p,q)$ implies simple causality, which is a stronger condition than strong causality (see \cite{MS}). But it is not possible to relax the condition of strong causality to just causality in Theorem \ref{teo1}, as Carter Spacetime proves (see Section \ref{carter}):
  
\begin{counter}\label{co}
Carter spacetime is causal and for each $p, q\in M$ the set $\overline{J(p,q)}$ is compact, but the spacetime is not globally hyperbolic. 
\end{counter}

There exists one last alternate definition of global hyperbolicity:
\begin{defn}\label{def4}
A spacetime $(M,g)$ is globally hyperbolic if and only if it is {\bf strongly causal} and for any metric $g'=\Omega g$, $\Omega>0$ conformal to the original one, the Lorentzian distance $d'$ associated to $g'$ is finite: $d'(p,q)<\infty, \forall p,q \in M$.
\end{defn}

With this background in mind, it is interesting to ask if, in last definition \ref{def4} and in definition \ref{def1}, it would be possible to relax the condition of strongly causal to just causal. As a matter of fact, it has been shown that since it is possible to do it in definition \ref{def3}, it is inmediately possible to do it in definition \ref{def1} (see \cite{MS}). For the other case, we tried to find a counterexample in Carter spacetime, because it seemed to be the best candidate due to the literature. Indeed, this spacetime is very simple and, since it is a foliation of compact manifolds, if the time separation or Lorentzian distance was finite for the metric $g$, then it would also be finite for every metric conformal to the original. However, as proven in Proposition \ref{proo}, Carter spacetime does not work as a counterexample. In any case, this result is interesting {\it per se} and yields some implications and ideas for alternate approaches described in the following section.

\section{Conclusion and open questions}\label{oq}
The importance of global hyperbolicity condition in General Relativity  has been justified in the previous section. Obviously, in order to have a better knowledge of this condition, it is required to understand the existing characterizations and work on them. Although the question we tried to solve remains open to our dissatisfaction, working on this problem has helped us understand better the issue and find possible alternate solutions of the problem. We recall here the open question:
\begin{oq}\label{oqs}
Is it possible, in Definition \ref{def4}, to relax the condition of strong causality to causality?
\end{oq}
The first attempt to answer this question was to find a counterexample, and we chose Carter spacetime as a candidate because of previous results about it.  We could continue working in that direction; the idea would be to find another causal but not strongly causal spacetime (maybe 3-dimensional, maybe not) with compact hypersurfaces $\{t=t_0\}$, such that if the distance is finite in the metric between every two points, it will stay finite in any conformal metric to the original. Carter spacetime with slight modifications could supply the desired counterexample. This approach could be the easiest, but it is necessary to avoid the curve $\b$ defined in Lemma \ref{lem1}, or to avoid the possibility of using the curve $\b$ to connect two points with a timelike curve of infinite length, always having in mind that the spacetime should remain causal and not become strongly causal in the process. If this idea does not work, maybe a new construction of a 3-dimensional spacetime would be necessary, which could be a harder task. Indeed, here again we need to be mindful of the fact that the structure of the causal cones should not give rise to a curve like $\b$ or the curves resulting from our construction. It could also happen that the wanted spacetime needs to be 4-dimensional or more, because lower dimensions would not work. Of course, searching for such counterexample and not finding it could imply that it is possible, in fact, to give a positive answer to  question \ref{oqs}. In both cases, a good understanding of the Theory of Causality and all the conditions in the causal ladder 
is required, as it will be argued below.

Let us review Definition \ref{def4}. On one hand, apart from the strong causality condition, a condition on the set of conformal metrics of the spacetime is imposed. This condition on the conformal metrics says that the finiteness of the distance is invariant under conformal changes of the metric, if the spacetime is globally hyperbolic.  One has to bear in mind that in a spacetime it is equivalent to study the causal behaviour of it and the conformal properties of the metric. This is because any spacetime $(M,g)$ and the manifold $M$ endowed with a  metric conformal to $g$, $(M,\Omega g)$ with $\Omega>0$, share the lightlike vectors at each point; consequently, both spacetimes have the same light cones at each point. Moreover, if $\nabla^*$ is the Levi-Civita connection associated to $(M,e^{2f} g)$, being $\Omega=e^{2f}$ the conformal factor of the metric, then:
$$\nabla^*_X Y=\nabla_X Y+X(f) Y+Y(f) X-g(X,Y)\nabla f$$
Observe that when taking $X=Y=\a'$ a lightlike tangent vector field associated to a lightlike geodesic $\a$ in $(M,g)$, one concludes that $\a$ is a lightlike pregeodesic in $(M,\Omega g)$, which can be reparametrized to a geodesic in that space. Therefore, the conformal structure of a spacetime determines the trajectories of photons and so, as said, it is equivalent to study the causal behaviour and the conformal structure of a spacetime.  On the other hand, it is important to bear in mind that the strong  causality condition causes the spacetime to have a very good behaviour in some cases. For example, in the class of strongly causal spacetimes, the Lorentzian distance $d$ determines the metric (see \cite[Th. 4.17]{BE}). Also, the Alexandrov topology, which basis is the intersection of the chronological  sets of a spacetime and is normally coarser than the topology of the manifold, coincides with the topology of the manifold in a strongly causal spacetime (\cite[pp. 196-197]{HE}).
Another example is that, since a strongly causal spacetime satisfies the future and past distinguishing condition, a point is uniquely determined by its chronological past or future, that is, $I^+(p)=I^+(q)$ or  $I^-(p)=I^-(q)$ if and only if $p=q$. Moreover, the limit curve of a sequence of curves in such spacetimes, if it exists, coincides with the convergence of curves in the $C^0$-topology (\cite[Proposition 3.34]{BE}). These are some examples of the good behaviour in a strongly causal spacetime. Therefore, it is important to make a careful study of how and when strong causality condition  is used in the proof of definition \ref{def4} viewed as a characterization of global hyperbolicity, and how could it be replaced by causal condition. 

Observe also that, while weakening the condition of strong causality in Definition \ref{def3}, it was proven that under the new conditions given in that definition, causality simplicity holds on the spacetime, and this is a stronger condition than strong causality. In the causal ladder there are four more causal conditions between strong causality and global hyperbolicity (see for example \cite{MS}). Thus, if we were able to prove that causal condition together with finite distance in any conformal metric to the given metric implies any of those four conditions, the desired result would follow. Nevertheless, there is also an alternate causal ladder, called {\it isocausality}, which  was intended to refine the standard causal ladder but resulted instead in a new hierarchy of spacetimes, with some elements in common with the old one (\cite{MS}, \cite{SG3}). Since the idea of isocausality was to compare spacetimes in the same standard causality level, this could perhaps lead us to a proof of our thesis. In this context, it would  be intereseting to find out if the finiteness of the Lorentzian distance is invariant under isocausality, which as far as we know, has not been studied until now. 

Summarizing, the results presented on Carter spacetime have opened the way to a better understanding of how one should approach question \ref{oqs}. On one hand there is the search of a counterexample. As argued, this counterexample could be either a modification of  Carter spacetime or a different space, but in any case  the resulting causal structure of the built spacetime should not allow the existence of a curve with the properties of $\b$ defined in Lemma \ref{lem1}, that could render an infinite Lorentzian distance between two given points, and should not be strongly causal. On the other hand, there remains the possibility of having a positive answer to our question.
For either path, some of the main points to consider would be the conformal structure of a spacetime, a deeper understanding of the behaviour of spacetimes that satisfy the strong causality condition, or any of the conditions in the causal ladder between that one and global hyperbolicity, and the study of the more recent isocausality ladder and its implications. In any case, the (positive or negative) answer to the thesis offered in our question would be of interest because it would provide a new point of view on global hyperbolicity, and a step forward in the understanding of the behaviour of globally hyperbolic spacetimes would be taken.

\end{document}